\documentclass[twocolumn,a4paper,eqsecnum,showkeys,showpacs,aps]{revtex4}
\usepackage{amsfonts,amssymb}
\usepackage{dcolumn}
\usepackage{amsmath}
\providecommand{\xii}{\xi\hspace{-1.5mm}
\raisebox{-1.5ex}{\textit{\scriptsize i}}\hspace{0.5mm}{}}
\providecommand{\xione}{\xi\hspace{-1.5mm}
\raisebox{-1.5ex}{{\scriptsize 1}}\hspace{0.2mm}{}}
\providecommand{\xitwo}{\xi\hspace{-1.5mm}
\raisebox{-1.5ex}{{\scriptsize 2}}\hspace{0.2mm}{}}
\providecommand{\xibartwo}{\bar{\xi}{}\hspace{-1.5mm}
\raisebox{-1.5ex}{{\scriptsize 2}}\hspace{0.2mm}{}}
\providecommand{\thetaj}{\theta\hspace{-2.0mm}
\raisebox{-1.5ex}{\textit{\scriptsize j}}\hspace{1.0mm}{}}
\providecommand{\thetaone}{\theta\hspace{-1.5mm}
\raisebox{-1.5ex}{{\scriptsize 1}}\hspace{0.2mm}{}}
\providecommand{\thetatwo}{\theta\hspace{-1.5mm}
\raisebox{-1.5ex}{{\scriptsize 2}}\hspace{0.2mm}{}}
\providecommand{\thetabartwo}{\bar{\theta}{}\hspace{-1.5mm}
\raisebox{-1.5ex}{{\scriptsize 2}}\hspace{0.2mm}{}}
\newfont{\cyr}{wncyr10 at 9pt}
\newfont{\cyrit}{wncyi10 at 9pt}
\begin{document}
\title{Adjoint representation of the graded Lie algebra
       \textit{osp}(2/1; $\mathbb{C}$) and its exponentiation}%
\author{Kostyantyn Ilyenko${}^*$}%
%\thanks{kost@ire.kharkov.ua}%
%\email{kost@ire.kharkov.ua}%
\affiliation{Institute for Radiophysics and Electronics, NAS of
             Ukraine, 12~Acad.~Proskura Street, Kharkiv 61085, Ukraine}%
\date{August 1, 2003}%
\begin{abstract}
\noindent We construct explicitly the grade star Hermitian adjoint
representation of \textit{osp}(2/1; $\mathbb{C}$) graded Lie
algebra. Its proper Lie subalgebra, the even part of the graded
Lie algebra \textit{osp}(2/1; $\mathbb{C}$), is given by
\textit{su}(2) compact Lie algebra. The Baker-Campbell-Hausdorff
formula is considered and reality conditions for the Grassman-odd
transformation parameters, which multiply the pair of odd
generators of the graded Lie algebra, are clarified.
\end{abstract}
               %Use showkeys class option if keyword display desired
\maketitle
%\tableofcontents
\section{Introduction}
\label{Intro}%
The number of gauge bosons of a gauge theory is given by the
number of generators of the underlying compact Lie algebra. The
requirement of being a \textit{compact} Lie algebra stems from the
reality of the action functional, which follows from the
properties of the Fermat principle in optics and the Feynman path
integral. This leads to physically admissible dynamics of a system
with such an action functional. In the case of a Yang-Mills theory
one utilizes properties of compact Lie algebras to ensure the real
property of the action functional. In this paper we explore how
one can utilize a $\mathbb{Z}_2$-graded extension of the compact
Lie algebra \textit{su}(2) for the purposes of defining a
meaningful gauge theory of the Yang-Mills type (see
\cite{Brooks1996,Ilyenko2001}). In particular, the matrices of the
graded Lie algebra generators in the adjoint representation and
the corresponding super-Killing form are explicitly calculated. By
exponentiating the graded algebra, we make preparations for
treating gauge transformations based on the graded Lie algebra and
study some properties of one-parameter families of transformations
of such a supergroup.
%-------------------------------------------------------------------------
\section{Graded Lie algebra $osp(2/1;~\mathbb{C})$}
\label{AlgGr}%
The algebra \textit{osp}(2/1;~$\mathbb{C}$) is a graded extension
of \textit{su}(2) algebra by a pair of odd generators, $\tau_{A}$,
which anticommute with one another and commute with the three even
generators, $T_{a}$, of \textit{su}(2). It is customary to assign
a degree, deg\,$T_\alpha$, to the even (deg\,$T_a$ = 0) and odd
(deg\,$\tau_A$ = 1) generators. We use the square brackets to
denote the commutator and the curly ones to denote the
anticommutator. The defining relations have the form,
\cite{Kac1977,Hughes1981,Brooks1996}:
\begin{eqnarray}
&[T_{a}, T_{b}] = i\varepsilon_{abc}T_{c},\;\;\; [T_{a}, \tau_{A}]
= \frac{1}{2}(\sigma_a)_{A}^{\hphantom{A}B}\tau_{B},& \nonumber \\
&\{\tau_{A}, \tau_{B}\} = \frac{i}{2}(\sigma^a)_{AB}T_{a}.&
\label{DefRel}
\end{eqnarray}
Summation is assumed over all repeated indices. Lowercase Roman
indices from the beginning of the alphabet run from 1 to 3;
uppercase Roman indices run over 1 and 2; $\delta_{ab}$ =
$\delta^{ab}$ ($\delta_{ab}$ = $\delta_{ba}$), $\varepsilon_{abc}$
($\varepsilon_{123}$ = $\varepsilon^{123}$ = 1) and
$\epsilon_{AB}$ ($\epsilon_{12}$ = $\epsilon^{12}$ = 1) are the
three dimensional identity matrix and the Levi-Civita totally
antisymmetric symbols in three and two dimensions, respectively;
the matrices $(\sigma_a)_{A}^{\hphantom{A}B}$ [$(\sigma^a)_{BA}$ =
$(\sigma^a)_{AB}$ = $\delta^{ab}(\sigma_b)_{AB}$ =
$\delta^{ab}(\sigma_b)_{A}^{\hphantom{A}C}\epsilon_{CB}$] are just
the usual Pauli matrices:% (the \textit{su}(2) soldering form,%
%\cite{Ashtekar87}):%
\begin{eqnarray}
&\!\!\!(\sigma^a)_{A}^{\hphantom{A}B} =
(\sigma_a)_{A}^{\hphantom{A}B} = \left[ \left(
\begin{array}{cc}
0 & 1 \\ 1 & 0
\end{array}
\right),
\left(
\begin{array}{rl}
 0 & i \\
-i & 0
\end{array} \right),%
\left( \begin{array}{cc} 1 &\! 0 \\ 0 &\! -1
\end{array}
\right) \right], & \nonumber \\%
%\;\;\;%
&\!\!\!(\sigma^a)_{AB} = (\sigma_a)_{AB} = \left[ \left(
\begin{array}{cc}
\!\! -1 & 0 \\
      0 & 1
\end{array}
\right), \left(
\begin{array}{rl}
-i &       0 \\
 0 & \!\! -i
\end{array} \right),%
\left( \begin{array}{cc}%
 0 &\! 1 \\%
 1 &\! 0
\end{array}
\right) \right]. & \nonumber %\\%
\label{PauliM}
\end{eqnarray}
We use the Levi-Civita symbols in two dimensions to raise and
lower uppercase Roman indices paying attention to their
antisymmetric properties:
\begin{displaymath}
\Sigma = \|\epsilon^{AB}\| = \left(
\begin{array}{rr}
0 &  1 \\ -1 & 0
\end{array}
\right) = \|\epsilon_{AB}\| = -\Sigma^{-1}.
\end{displaymath}
Note that,  as concerned to these indices, we are
working with two-component spinors and adopt conventions of the
book~\cite{Stewart}. We shall follow those conventions as more
suitable for our purposes even when complex
conjugation of spinor and Grassman quantities is involved.\\%
\indent In the adjoint representation the matrices $T_a$ and
$\tau_A$ can be written as follows (solid lines are drawn to
emphasize their block structure):
\begin{widetext}
\begin{displaymath}
 T_1 = \left( \mbox{%
\begin{tabular}{ccc|cc}
 $0$&$0$&$i$& $0$&$0$ \\
 $0$&$0$&$0$& $0$&$0$ \\
$-i$&$0$&$0$& $0$&$0$ \\ \hline
 $0$&$0$&$0$& $0$&$1/2$ \\
 $0$&$0$&$0$& $1/2$&$0$ \\
\end{tabular} }
              \right),\;\;\;
T_2 = \left(\mbox{%
\begin{tabular}{ccc|cc}
$0$&$-i$&$0$& $0$&$0$ \\
 $i$&$0$&$0$& $0$&$0$ \\
 $0$&$0$&$0$& $0$&$0$ \\ \hline
 $0$&$0$&$0$& $0$&$-i/2$ \\
 $0$&$0$&$0$& $i/2$&$0$ \\
\end{tabular} }
              \right),\;\;\;
T_3 = \left(\mbox{%
\begin{tabular}{ccc|cc}
 $0$&$0$&$0$& $0$&$0$ \\
$0$&$0$&$-i$& $0$&$0$ \\
 $0$&$i$&$0$& $0$&$0$ \\ \hline
 $0$&$0$&$0$& $1/2$&$0$ \\
 $0$&$0$&$0$& $0$&$-1/2$ \\
\end{tabular} }
             \right),
\end{displaymath}
%\vspace{}%
${}^*$Electronic address: kost@ire.kharkov.ua%
\pagebreak
\begin{equation}
 T_4 \equiv \tau_1 = \frac{1}{2}\left(\mbox{%
\begin{tabular}{ccc|cc}
 $0$&$0$&$0$& $0$&$1$ \\
 $0$&$0$&$0$& $-1$&$0$ \\
 $0$&$0$&$0$& $-i$&$0$ \\ \hline
$-i$&$0$&$0$& $0$&$0$ \\
$0$&$-i$&$1$& $0$&$0$ \\
\end{tabular} }
              \right),\;\;\;
  T_5 \equiv \tau_2 = \frac{1}{2}\left(\mbox{%
\begin{tabular}{ccc|cc}
  $0$&$0$&$0$& $1$&$0$ \\
  $0$&$0$&$0$& $0$&$1$ \\
  $0$&$0$&$0$& $0$&$-i$ \\ \hline
$0$&$-i$&$-1$& $0$&$0$ \\
  $i$&$0$&$0$& $0$&$0$ \\
\end{tabular} }
              \right).
\label{AdjRepM}
\end{equation}
\end{widetext}
Let us denote $T_4$ = $\tau_1$, $T_5$ = $\tau_2$ and employ
lowercase Greek indices from the beginning of the alphabet
($\alpha$, $\beta$, %\\ \\%
etc.) to run over the whole set, $T_\alpha$, of the generators of
\textit{osp}(2/1;~$\mathbb{C}$). We then find that the
non-degenerate super-Killing form, $B(T_\alpha, T_\beta)$, is
given by
\begin{equation}
B(T_\alpha, T_\beta) = \frac{2}{3}\,\mbox{str}(T_\alpha T_\beta) =
\left(\mbox{%
\begin{tabular}{c|c}
$\delta_{ab}$& $0$ \\ \hline
          $0$& $i\epsilon_{AB}$ \\
\end{tabular}}
%\begin{array}{cc}
% \delta_{ab} &        0      \\
%     0       & i\epsilon_{AB}
%\end{array}%
 \right),
\label{sKillingF}
\end{equation}
where the supertrace operation is adopted from
\cite[pp.~18-19,~42]{Cornwell1989}. It is a linear operation,
which in a certain way inherits all the properties of the ordinary
trace.

It turns out that all of the generators are grade star Hermitian
(the operation is denoted by ${}^\ddag$): on the even generators
the operation coincides with ordinary Hermitian conjugation
(${}^+$) while the odd ones obey more complicated relations.
Following the papers~\cite{SNR19771,SNR19772}, we shall call them
the grade star Hermiticity conditions: $\tau_{\pm}^\ddagger$ =
$\pm\tau_{\mp}$ ($\tau_{\pm}$ = $\tau_1$ $\pm$ $i\tau_2$).

Let us consider complex-valued matrices divided into blocks
according to the scheme (see (\ref{AdjRepM}) and
(\ref{sKillingF})):
\begin{eqnarray}
M_{even} = \left(\mbox{{\small%
\begin{tabular}{c|c}
$A$& $0$ \\ \hline
$0$& $D$ \\
\end{tabular}}} \right) &
\mbox{and} &
% \begin{array}{*2{c}}
%    A & 0 \\
%    0 & D \\
%    \end{array} \right) &
M_{odd} = \left(\mbox{{\small%
\begin{tabular}{c|c}
$0$& $B$ \\ \hline
$C$& $0$ \\
\end{tabular}}} \right), \nonumber
%\begin{array}{*2{c}}
%    0 & B \\
%    C & 0 \\
%    \end{array} \right|. \nonumber
\end{eqnarray}
where $B$ and $C$ are 2$\times$3 rectangular blocks and $A$ and
$D$ are 3$\times$3 and 2$\times$2 square blocks, respectively. On
these matrices the supertrace operation is defined by
str$M_{even}$~= tr$A$ - tr$D$ and str$M_{odd}$ = 0 (here ``tr''
denotes the ordinary trace), while the grade star Hermiticity
condition reads
\begin{eqnarray}
M_{even}^\ddag = \left(\mbox{{\small%
\begin{tabular}{c|c}
$A^+$& $0$ \\ \hline
$0$& $D^+$ \\
\end{tabular}}} \right) &
\mbox{and} &
M_{odd}^\ddag = \left(\mbox{{\small%
\begin{tabular}{c|c}
$0$& $-C^+$ \\ \hline
$B^+$& $0$  \\
\end{tabular}}} \right). \nonumber
\end{eqnarray}

We shall also use multiplication of algebra generators by scalars.
Such an operation must take into account that Grassman-odd scalars
anticommute with the odd algebra generators while commute with
complex numbers and the even algebra generators,
\cite{Gendeshtein-Krive}. The following construction possesses all
of these properties. Let $a$ be a scalar and deg\,$a$ be its
degree (0 or 1 depending on whether it is Grassman-even or
Grassman-odd, respectively). Then multiplication by $a$ is defined
as follows:
\begin{widetext}
\begin{eqnarray}
aM_{odd}\hspace{0.5em} =
\left(\mbox{{\small%
\begin{tabular}{c|c}
$a$& $0$ \\ \hline
$0$& $(-1)^{\mbox{deg\,\textit{a}}}a$ \\
\end{tabular}}} \right)
\left(\mbox{{\small%
\begin{tabular}{c|c}
$0$& $B$ \\ \hline
$C$& $0$ \\
\end{tabular}}} \right)%
& = & (-1)^{\mbox{deg\,\textit{a}}}
\left(\mbox{{\small%
\begin{tabular}{c|c}
$0$& $B$ \\ \hline
$C$& $0$ \\
\end{tabular}}} \right)
\left(\mbox{{\small%
\begin{tabular}{c|c}
$a$& $0$ \\ \hline
$0$& $(-1)^{\mbox{deg\,\textit{a}}}a$ \\
\end{tabular}}} \right)
=(-1)^{\mbox{deg\,\textit{a}}}M_{odd}a,  \nonumber \\
 aM_{even} =
\left(\mbox{{\small%
\begin{tabular}{c|c}
$a$& $0$ \\ \hline
$0$& $(-1)^{\mbox{deg\,\textit{a}}}a$ \\
\end{tabular}}} \right)
\left(\mbox{{\small%
\begin{tabular}{c|c}
$A$& $0$ \\ \hline
$0$& $D$ \\
\end{tabular}}} \right)%
& = &
\left(\mbox{{\small%
\begin{tabular}{c|c}
$A$& $0$ \\ \hline
$0$& $D$ \\
\end{tabular}}} \right)
\left(\mbox{{\small%
\begin{tabular}{c|c}
$a$& $0$ \\ \hline
$0$& $(-1)^{\mbox{deg\,\textit{a}}}a$ \\
\end{tabular}}} \right)
=M_{even}a. \nonumber
\end{eqnarray}%
\end{widetext}
%-------------------------------------------------------------------------
\section{Baker-Campbell-Hausdorff formula}%
\label{BCH-f}%
Given a Lie algebra one can turn over to a Lie group by
exponentiating the generators. In the case of a graded Lie algebra
we are faced with a problem: anticommutators seem to rule out the
application of the Baker-Campbell-Hausdorff formula, which is
necessary to prove that subsequent transformations do not leave
the group manifold. This problem is solved via introduction of
Grassman-odd parameters (cf.,~\cite{Berezin1970}). In the case
under consideration these are Grassman-odd \textit{su}(2)-spinors
$\xi^A$, $\theta^A$, etc., which multiply the odd generators. They
are included on equal footing with ordinary (Grassman-even)
parameters $\varepsilon^a$ multiplying the even generators. By
definition, $\xi^A$, $\theta^A$, etc. satisfy
\begin{equation*}
[\varepsilon^a,\,\theta^A]=0,\,\,\{\xi^A,\,\xi^B\}=
\{\theta^A,\,\theta^B\}=0,\,\,\{\xi^A,\,\theta^B\}=0.
\end{equation*}
Then, the graded Lie algebra can be written only in terms of
commutators:
\begin{equation*}
[\xi^A\tau_A,\,\theta^B\tau_B]=-\frac{i}{2}(\xi^{\{A}\theta^{B\}}
+ \xi^{[A}\theta^{B]})(\sigma^a)_{AB}T_a,%
\end{equation*}
where $\xi^{\{A}\theta^{B\}} = 1/2(\xi^A\theta^B + \xi^B\theta^A)$
and $\xi^{[A}\theta^{B]} = 1/2(\xi^A\theta^B - \xi^B\theta^A)$ are
convenient shorthand notations. This result was obtained using
anticommutator for odd generators in definition (\ref{DefRel}).
Using a fundamental fact of spinor algebra,
$\epsilon_{AB}\epsilon_{CD}$ $+$ $\epsilon_{AC}\epsilon_{DB}$ $+$
$\epsilon_{AD}\epsilon_{BC}$ $=$ $0$, one can calculate
\begin{equation*}
\xi^{[A}\theta^{B]}=\frac{1}{2}(\xi_C\theta^C)\epsilon^{AB}.
\end{equation*}
From symmetry of $(\sigma^a)_{AB}$ in the uppercase indices, it
then follows that
\begin{eqnarray}
[\xi^A\tau_A,\,\theta^B\tau_B]%
&    =   &%
-\frac{i}{2}\xi^{\{A}\theta^{B\}}(\sigma^a)_{AB}T_a \nonumber \\%
& \equiv &%
-\frac{i}{2}[\xi^{A},\theta^{B}](\sigma^a)_{AB}T_a%
\label{ComOdd}
\end{eqnarray}
and, in particular, the commutator
$[\theta^A\tau_A,\,\theta^B\tau_B]$ vanishes identically. One can
also calculate
\begin{eqnarray}
\left[\kappa^aT_a,\,\varepsilon^bT_b\right] & = &
i\kappa^{[a}\varepsilon^{b]}\varepsilon_{abc}T_c, \nonumber \\
\left[\varepsilon^{a}T_{a},\,\theta^{A}\tau_{A}\right] & = &
\tilde{\theta}{}^{B}\tau_{B}, \label{ComEve}
\end{eqnarray}
where $\tilde{\theta}{}^B$ $=$
$1/2\varepsilon^a$$\theta^A$$(\sigma_a)_A^{\hphantom{A}B}$ is
again a Grassman-odd transformation parameter.

Group elements are obtained by exponentiating the algebra
\begin{equation}
U(\varepsilon,\,\theta) = \exp[i(\varepsilon^aT_a + \theta^A\tau_A)]%
\label{SPT}
\end{equation}
and the Baker-Campbell-Hausdorff formula,
\begin{equation*}
\exp(M)\exp(N) = \exp(M + N + \frac{1}{2}[M,\,N] + \hdots),
\end{equation*}
may be applied to determine motion in the parameter space under a
(left) multiplication with a group element $U(\kappa,\,\xi)$:
\begin{equation*}
U(\varepsilon' ,\,\theta') =
U(\kappa,\,\xi)U(\varepsilon,\,\theta).
\end{equation*}
Substituting (\ref{SPT}) and using (\ref{ComOdd}), we obtain after
some algebra
\begin{eqnarray}
\varepsilon'{}^a & = & \varepsilon^a + \kappa^a +
\frac{1}{2}\varepsilon_b\kappa_c\,\varepsilon^{bca} +
\frac{1}{4}[\xi^{A},\theta^{B}](\sigma^a)_{AB} + \ldots \nonumber \\
\theta'{}^A & = & \theta^A + \xi^A + \frac{i}{4}(\kappa_b\theta^B
- \varepsilon_b\xi^B)(\sigma^b)_{B}^{\hphantom{B}A} + \ldots%
\label{PTrans}
\end{eqnarray}
Here the dots denote the remaining contribution from linear
combinations of $k$-fold ($k > 2$, $k \in \mathbb{Z}$) commutators
of $M$ and $N$ to the Baker-Campbell-Hausdorff formula.

The last term in the first equation of system (\ref{PTrans}) needs
to be investigated in more detail. First, let us calculate that
\begin{eqnarray}
  2[\xi^A,\theta^B](\sigma^a)_{AB} & = &
  \xi_A\epsilon^{AB}(\sigma^a)_{B}^{\hphantom{B}C}\theta_C -
  \theta_A\epsilon^{AB}(\sigma^a)_{B}^{\hphantom{B}C}\xi_C \nonumber \\
  & = & \xi^T\Sigma\sigma^a\theta - \theta^T\Sigma\sigma^a\xi,
\label{CVect}
\end{eqnarray}
where we employed some self-evident matrix notations. Comparing
the result (\ref{CVect}) and a description of
\textit{su}(2)-spinors of 3D~Euclidean space in the book
\cite[p.~48]{Cartan1966}, one immediately realizes that the last
term of the first equation in system (\ref{PTrans}) is, in
general, a \textit{complex} vector of 3D~Euclidean space, e.g. it
transforms like a vector under \textit{SO}(3) transformations.
Second, the representation (\ref{CVect}) tells us that components
of this vector vanish if $\xi_A$=$\theta_A$ as required by a
property of a one-parameter subgroup of transformations
(\ref{SPT}). Finally, this vector also has all components equal to
zero if $\xi_A$ = $-\,\theta_A$. This shows that the inverse of
the group element $U(\varepsilon,\,\theta)$ has the form
\begin{equation}
U^{-1}(\varepsilon,\,\theta) = \exp[-i(\varepsilon^aT_a + \theta^A\tau_A)].%
\label{ISPT}
\end{equation}

If one intends, as we actually do, to treat $\varepsilon^a$,
$\kappa^a$, etc. as real-valued transformation parameters, then it
is necessary to impose some conditions on the
\textit{su}(2)-spinors $\xi_A$, $\theta_A$, etc. in order to
ensure that (\ref{CVect}) will be a \textit{real} 3D~Euclidean
vector. Such a condition must be compatible with transformation
properties of the corresponding space of \textit{su}(2)-spinors,
$\xi_A$, and take into account that its members are also
Grassman-odd quantities. In fact, this condition should involve a
passage from an \textit{su}(2)-spinor to its conjugate and, thus,
rely on the definition of an anti-involution in the space of
spinors (see, e.g. \cite[p.~100]{Cartan1966}). Let us observe
first that for a Grassman algebra on one generator the last term
in the first relation in (\ref{PTrans}) vanishes identically. This
is a somewhat trivial situation. The next non-trivial one arrises
when all \textit{su}(2)-spinors under consideration take values in
a Grassman algebra on two odd generators, $\varepsilon_1$ and
$\varepsilon_2$: $\varepsilon_1^2$ = $\varepsilon_2^2$ = 0,
$\varepsilon_1\varepsilon_2$ = $-\varepsilon_2$$\varepsilon_1$
(see, e.g. \cite[p.~7]{Cornwell1989}). It is hoped that there
would be no misunderstanding due to the fact that we use the same
kernel letter to denote vector components $\varepsilon^a$ and the
two odd generating elements of the Grassman algebra. We shall
employ lowercase Roman indices from the middle of the alphabet
running over 1 and 2 to enumerate the decompositions of various
quantities in the corresponding basis of the Grassman algebra.
Decomposing $\xi_A$ and $\theta_A$ into this basis one obtains
\begin{displaymath}
\xi_A = \xii{}_A\varepsilon_i\,\,\,\mbox{and} \,\,\,\theta_B =
\thetaj{}_B\varepsilon_j,
\end{displaymath}
%$$\xii, \xij, \thetai, \thetaj, \xione, \xitwo, \thetaone, \thetatwo$$%
where $\xii{}_A$ and $\thetaj{}_B$ are ordinary, i.e. commuting,
\textit{su}(2)-spinors of 3D~Eclidean space, and summation over
repeated indices is assumed. In this case we can write
\begin{eqnarray}\label{CVectG}
\frac{1}{2}(\xi^A\theta^B - \theta^A\xi^B)(\sigma^a)_{AB} & = & \\
& = & \varepsilon_1\varepsilon_2(\xione^T\Sigma\sigma^a\thetatwo -
\thetaone^T\Sigma\sigma^a\xitwo). \nonumber
\end{eqnarray}
Now we shall impose some additional conditions on
\textit{su}(2)-spinors $\xii{}_A$, $\thetaj{}_A$, etc. to ensure
that (\ref{CVectG}) gives a \textit{real} Grassman-even
3D~Euclidean vector. One way of doing so in a manner preserving
all the spinor transformations properties is to define
\begin{equation}\label{CC}
    \xione{}_A = iC_{A}^{\hphantom{A}B'}\xibartwo{}_{B'}, \,\,\,
    \thetaone{}_A = iC_{A}^{\hphantom{A}B'}\thetabartwo{}_{B'}, \,\,\,
    \mbox{etc.},
\end{equation}
where the `charge conjugation' matrix $C$ ($C\overline{C}$ =
${}-I$) is given by
\begin{displaymath}
    C = \|C_{A}^{\hphantom{A}B'}\| =
\left( \begin{array}{cc}
      0 & 1 \\
\!\! -1 & 0
\end{array} \right)
= \|\overline{C}_{A'}^{\hphantom{A'}B}\| = \overline{C}.
\end{displaymath}
In (\ref{CC}) a bar over the spinors in the left-hand sides of the
relations and primes over the indices denote complex conjugation.
We again adhere to Penrose notations when spinors are conserned,
\cite{Stewart}. The charge conjugation matrix,
$C_A^{\hphantom{A}B'}$, is responsible for invariant preservation
of spinor properties (for details see, e.g. the review article
\cite[pp.~108~--~109]{Rashevskii1955}, where this object is
denoted by $\stackrel{*}{\Pi}{}^\lambda_{\dot{\mu}}$; also compare
with the treatment in \cite[p.~100]{Cartan1966}). As seen from the
relation (\ref{CC}) in this case each Grassman-odd
\textit{su}(2)-spinor $\xi_A$, $\theta_A$, etc. is defined by a
single ordinary (Grassman-even) \textit{su}(2)-spinor. For the
sake of notations denoting, respectively,
\begin{displaymath}
\xitwo{}_A = \eta_A\,\,\,\mbox{and}\,\,\,\thetatwo{}_B =
\vartheta_B,
\end{displaymath}
we write
\begin{eqnarray}\label{RVect}
\varkappa^a & \equiv & \xione^T\Sigma\sigma^a\thetatwo
- \thetaone^T\Sigma\sigma^a\xitwo = \nonumber \\
& = & i(\bar{\eta}{}^TC{}^T\Sigma\sigma^a\vartheta -
\bar{\vartheta}{}^TC{}^T\Sigma\sigma^a\eta).
\end{eqnarray}
On comparison with \cite[p.~50]{Cartan1966}, one can check that
$\varkappa^a$ is indeed a \textit{real} 3D~Euclidean vector. In
components
it reads:%
%\pagebreak%
\begin{eqnarray}\label{RVectComps}
\varkappa^1 & = & i(\bar{\eta}{}_{1'}\vartheta_2 -
\bar{\vartheta}{}_{2'}\eta_1 + \bar{\eta}{}_{2'}\vartheta_1 -
\bar{\vartheta}{}_{1'}\eta_2), \nonumber \\
  \varkappa^2 & = & \hphantom{i(}\bar{\eta}{}_{2'}\vartheta_1 +
  \bar{\vartheta}{}_{1'}\eta_2 - \bar{\eta}{}_{1'}\vartheta_2 -
  \bar{\vartheta}{}_{2'}\eta_1, \\
  \varkappa^3 & = & i(\bar{\eta}{}_{1'}\vartheta_1 - \bar{\vartheta}{}_{1'}\eta_1
  - \bar{\eta}{}_{2'}\vartheta_2 + \bar{\vartheta}{}_{2'}\eta_2). \nonumber
\end{eqnarray}
These are obviously real quantities and the vector $\varkappa^a$
vanishes if and only if $\eta_A$ = $\pm\vartheta_A$ as required.
%-------------------------------------------------------------------------
\section{Discussion}
%\indent%
To the best of the author's knowledge the graded extension of
\textit{u}(1) Lie algebra for the first time was considered by
Berezin and Kac in \cite{Berezin1970}. However, they didn't build
the adjoint representation, neither they calculated the
super-Killing form no studied the conditions of reality for a
product of two one-parameter families of transformations. From the
current standpoint, the last task could be achieved in an
analogous manner to that of presented in Sec.~\ref{BCH-f}. The
only necessary remark here is that one would use complex Euclidean
spinors in one dimension, i.e. complex numbers, and conditions of
(\ref{CC})-type. In the present paper we treated the next simplest
case: the graded extension of \textit{su}(2) Lie algebra.

For a reader's convenience we present below a table with some
first members of the two major series of graded Lie algebras (cf.,
\cite{FK1976}). Their respective proper Lie subalgebras and the
numbers of even, $m$, and odd, $n$, generators are shown,
\cite{Kac1977,Cornwell1989}. We are concerned here only with
compact choice for the proper Lie subalgebras of the graded Lie
algebras:
\begin{widetext}
\begin{center}
\begin{tabular}{|c|c|c|c|c|c|}\hline
 \multicolumn{6}{|c|}{\rule[-1.2ex]{0ex}{4.5ex}\bfseries Low-dimensional
 graded Lie algebras of two main series}\\ \hline
\rule[0ex]{0ex}{2.8ex}
 & proper Lie subalgebra & \textit{u}(1) & \textit{u}(1)$\oplus$\textit{su}(2)
 & \textit{u}(1)$\oplus$\textit{su}(2)$\oplus$\textit{su}(2) &
 \textit{u}(1)$\oplus$\textit{su}(3)\\ \cline{2-6}
\rule[0ex]{0ex}{2.8ex}
 \raisebox{1.9ex}[0pt]{\textit{spl}($p/q$; $\mathbb{C}$); $p, q > 0$:}
 & $m$ -- dim. even subsp. & 1 & 4 & 7 & 9 \\ \cline{2-6}
\rule[0ex]{0ex}{2.8ex}
 \raisebox{1.7ex}[0pt]{$m = p^2 + q^2 - 1, n = 2pq$} & $n$ \,-- dim.
 \,odd\, subsp. & 2 & 4 & 6 & 6 \\
 \hline\hline
\rule[0ex]{0ex}{2.8ex}
  & $m$ -- dim. even subsp. & 3 & 4 & 6 & 9 \\ \cline{2-6}
\rule[0ex]{0ex}{2.8ex}
 \raisebox{2.9ex}[0pt]{\textit{osp}($p/q$; $\mathbb{C}$); $p \geq 1$,
 even $q > 0$:} & $n$ \,-- dim. \,odd\, subsp. & 2 & 4 & 6 & 8 \\ \cline{2-6}
\rule[0ex]{0ex}{2.8ex}
 \raisebox{1.3ex}[0pt]{$m = \dfrac{p(p-1)}{2} + \dfrac{q(q+1)}{2}, n = pq$}
 & proper Lie subalgebra & \textit{su}(2) &
 \textit{u}(1)$\oplus$\textit{su}(2)
 & \textit{su}(2)$\oplus$\textit{su}(2) &
 \textit{su}(2)$\oplus$\textit{su}(2)$\oplus$\textit{su}(2)\\
 \hline
\end{tabular}
\end{center}
\vspace{-0.7ex}
\end{widetext}
%-------------------------------------------------------------------------
\section*{Acknowledgements}
%\indent%
I am grateful to Dr.~T.S.~Tsou for interest in this work and to
Prof.~Yu.P.~Stepanovsky for numerous helpful  discussions. I would
also like to acknowledge an inspiring criticism from
Dr.~V.~Pidstrigach at an early stage of this development.
%----------------------------------------------------------------------------

%---------------------------------------------------------------------------

\begin{thebibliography}{99}
\bibitem{Brooks1996} R.~Brooks, A.~Lue, ``The monopole equations
  in topological Yang-Mills'', {\em J.~Math.~Phys.} {\bf 37}, 1100 --
  1105 (1996); Preprint hep-th/9412206.
\bibitem{Ilyenko2001} K.~Ilyenko, ``Field strength for graded Yang-Mills
  theory'', {\em Problems Atom. Sci. Tech.} {\bf 6}(1), 74 --
  75 (2001); Preprint hep-th/0307230.
\bibitem{Kac1977} V.~Kac, ``Representations of classical Lie
  superalgebras'', in: {\it Differential Geometrical Methods in
  Mathematical Physics II}, Eds. K.~Bleuler, H.R.~Petry and
  A.~Reetz, LMS Lecture Notes, Springer-Verlag, Berlin, {\bf 676}, 597
  -- 626 (1978).
\bibitem{Hughes1981} J.W.~Hughes, ``Representations of
  \textit{osp}(2,1) and the metaplectic representation'',
  {\em J.~Math.~Phys.} {\bf 22}, 245 -- 250 (1981).
\bibitem{Stewart} J.~Stewart, ``Advanced general relativity'',
  Cambridge University Press, Cambridge 1996.
\bibitem{Cornwell1989} J.F.~Cornwell, ``Group theory in physics'', Vol.~3,
  Mac\-Mil\-lan Publishers, London 1989.
\bibitem{SNR19771}  M.~Scheunert, W.~Nahm, Y.~Rittenberg, ``Graded Lie
  algebras: generalization of Hermitian representations'', {\em J.~Math.~Phys.}
  {\bf 18}, 146 -- 154 (1977).
\bibitem{SNR19772}  M.~Scheunert, W.~Nahm, Y.~Rittenberg, ``Irreducible
  representations of the \textit{osp}(2,1) and \textit{spl}(2,1) graded Lie
  algebras'', {\em J.~Math.~Phys.} {\bf 18}, 155 -- 162 (1977).
\bibitem{Gendeshtein-Krive} {\cyr L.E1.~Gendeshte}$\check{\mbox{\cyr i}}${\cyr
  n, I.V.~Krive}, ``{\cyr Su\-per\-sim\-met\-riya v kvan\-to\-vo}$\check{\mbox{\cyr i}}$
  {\cyr me\-kha\-ni\-ke}'', {\cyrit UFN} {\bf 146}(4), 553 -- 590 (1985) [Eng.
  transl.: L.E.~Gendenshtein, I.V.~Krive, ``Supersymmetry in quantum mechanics'',
  \textit{Sov. Phys.~--~Usp.} {\bf 28}(8), 645 -- 666 (1985)].
\bibitem{Berezin1970} {\cyr F.A.~Berezin, G.I.~Kats},
  ``{\cyr Gruppy Li s kom\-mu\-ti\-ruyu\-wi\-mi i
  an\-ti\-kom\-mu\-ti\-ruyu\-wi\-mi parametrami}'',
  {\cyrit Mat. sb.} {\bf 82}(3), 343 -- 359 (1970) [Eng. transl.:
  F.A.~Be\-re\-zin, G.I.~Kac, ``Lie groups with commuting and anticommuting
  parameters'', \textit{Math. USSR~--~Sb.} {\bf 11}, 311 -- 325 (1971)].
\bibitem{Cartan1966} E.~Cartan, ``The theory of spinors'', Hermann
  Publishers, Paris 1966.
\bibitem{Rashevskii1955} {\cyr P.K.~Rashevski}$\check{\mbox{\cyr
  i}}$, ``{\cyr Teo\-riya spi\-no\-rov}'', {\cyrit UMN} {\bf 10}(2), 3 --
  110 (1955) [Eng. transl.: P.K.~Rashevskij, ``The theory of
  spinors'', {\em  Transl. Am. Math. Soc.}~{\bf II} (Ser.~6), 1 -- 110 (1957)].
\bibitem{FK1976}  P.G.O.~Freund, I.~Kaplansky, ``Simple supersymmetries'',
  {\em J.~Math.~Phys.} {\bf 17}, 228 -- 231 (1976).
\end{thebibliography}
\end{document}